\documentclass[preprint,12pt]{aastex6}

\usepackage{amsmath}

\begin{document}
\title{Spectral Diagnostics of Solar Photospheric Bright Points}

\author{Q. Hao\altaffilmark{1,2}, C. Fang\altaffilmark{1,2}, M. D. Ding\altaffilmark{1,2},
        Z. Li\altaffilmark{1,2}, and W. Cao\altaffilmark{3,4}}

\altaffiltext{1}{School of Astronomy and Space Science, Nanjing University, Nanjing 210023, China}
\altaffiltext{2}{Key Laboratory of Modern Astronomy and Astrophysics (Nanjing University), Ministry of Education, China}
\altaffiltext{3}{Big Bear Solar Observatory, New Jersey Institute of Technology,~40386 North Shore Lane, Big Bear City, CA 92314, USA}
\altaffiltext{4}{Center for Solar-Terrestrial Research, New Jersey Institute of Technology, University Heights, Newark, NJ 07102, USA\\}

\email{fangc@nju.edu.cn}
\email{haoqi@nju.edu.cn}

\begin{abstract}
By use of the high-resolution spectral data and the broadband imaging obtained with the Goode Solar Telescope at the Big Bear Solar Observatory on 2013 June 6, the spectra of three typical photospheric bright points (PBPs) have been analyzed. Based on the H$\alpha$ and \ion{Ca}{2} 8542 {\AA} line profiles, as well as the TiO continuum emission, for the first time, the non-LTE semi-empirical atmospheric models for the PBPs are computed. The attractive characteristic is the temperature enhancement in the lower photosphere. The temperature enhancement is about 200 -- 500 K at the same column mass density as in the atmospheric model of the quiet-Sun. The total excess radiative energy of a typical PBP is estimated to be 1$\times$10$^{27}$ - 2$\times$10$^{27}$ ergs, which can be regarded as the lower limit energy of the PBPs. The radiation flux in the visible continuum for the PBPs is about 5.5$\times$10$^{10}$ ergs cm$^{-2}$ s$^{-1}$. Our result also indicates that the temperature in the atmosphere above PBPs is close to that of a plage. It gives a clear evidence that PBPs may contribute significantly to the heating of the plage atmosphere. Using our semi-empirical atmospheric models, we estimate self-consistently the average magnetic flux density $B$ in the PBPs. It is shown that the maximum value is about one kilo-Gauss, and it decreases towards both higher and lower layers, reminding us of the structure of a flux tube between photospheric granules.
\end{abstract}
\keywords{ Sun: atmosphere -- Sun: photosphere -- Sun: chromosphere}

\maketitle

\section{Introduction}

Photospheric Bright Points (PBPs) are small bright features observed in the photosphere. They exhibit bright emission in the visible continuum, such as the G-band and TiO band, or in the far wings of some spectral lines, such as H$\alpha$ and \ion{Ca}{2} H/K lines. Generally, PBPs appear in the intergranular lanes and correspond to strong magnetic flux concentrations \citep{Spruit1976,Schu1988,Solanki1993,Steiner2001,Steiner2007,Riethmuller2014}. Thus, the study of the physical characteristics of PBPs, especially their temperature and density stratifications, can help understand the properties and  physical mechanisms occurring in small strong magnetic flux concentration, which may in turn reveal a possible source of coronal heating~\citep{Withbroe1977,Srivastava2017}.

PBPs have been studied since the 1970s. It was found that the PBPs are best shown at H$\alpha$+2 {\AA}, and in the continuum images taken at 6439 {\AA} ~\citep{Dunn1973}. In the images taken with the broadband filters at \ion{Ca}{2} K and H$\alpha$ bands, PBPs occur singly or appear in chains or "crinkles", lying in the intergranular lanes ~\citep{Mehltretter1974} . Since then, several observations of PBPs have been reported in the literature, together with theoretical models based on the concept of flux tube (e.g.,~\cite{Spruit1976,Schu1988,Steiner2007}). Similarly, semi-empirical models based on observations of Stokes I and/or V profiles, as well as continuum contrasts (especially center-to-limb variation) have been put forward~\citep{Solanki1993,Lagg2010,Cristaldi2017}. Most of them are filling factor dependent and are derived assuming Local Thermodynamic Equilibrium (LTE) (see ~\cite{Solanki1993,Steiner2007} for review). Later on, many authors used G-band images, combined with the broadband H$\alpha$ and/or \ion{Ca}{2} (H, K) imaging observations \citep{Berger1995,Berger1996,Berger1998,Sainz2011,Chitta2012,Xiong2017}, while some authors used the broadband imaging at TiO 7057 {\AA} ~\citep{Cao2010,LiuZ2014,Liu2018} to study PBPs. Recently, owing to the development of adaptive optics and image processing technology, high-spatial resolution observations have been used to explore the characteristics of PBPs in detail. For example, a 3D track-while-detect (TWD) method was used to detect and track 27,696 PBPs in the images of G-band and \ion{Ca}{2} H line from SOT/Hinode ~\citep{Xiong2017}. A new algorithm was also developed to identify and track 2010 PBPs in TiO 7057 {\AA} images observed by the New Vacuum Solar Telescope of the Yunnan Observatory ~\citep{Liu2018} . They classified PBPs into isolated (individual) and non-isolated (multiple PBPs displaying splitting and merging behaviors) sets. Based on statistical studies and detailed analysis of individual cases, a series of characteristics of PBPs have been obtained. That is, the diameters of PBPs range from 150 -- 400 km ~\citep{Berger1995,Xiong2017,Liu2018}; the lifetimes of PBPs are about 200 -- 1000 s or more ~\citep{Muller1983,deWijn2005,Jafarzadeh2013,Keys2014,Xiong2017,Liu2018}, and the brightness varies from 0.8 to 1.8 times the average background intensity ~\citep{Mehltretter1974,Berger1995,Liu2018}. PBPs show random-walk characteristics, with an average speed of 0.5 -- 3 km s$^{-1}$ ~\citep{Berger1996,Nisenson2003,Criscuoli2012,Chitta2012,Keys2014}. As several authors indicated, the physical properties deduced from observations highly depend on the resolution of the observations (see, e.g.,~\cite{Spruit1981,Knoelker1988,Criscuoli2009}), while so far the high-spectral resolution observations of PBPs in both H$\alpha$ and \ion{Ca}{2} 8542 {\AA} lines are rare. It is quite expected to have the high-spectral and high-spatial resolution observations and make detailed analysis.

In this paper, we analyze the spectra of H$\alpha$ and \ion{Ca}{2} 8542 {\AA} lines and the broadband TiO images of three typical PBPs, which were observed with the 1.6-meter Goode Solar Telescope (GST) ~\citep{Cao2010,Goode2012} at Big Bear Solar Observatory (BBSO). These spectral and imaging data were obtained simultaneously with high-spatial resolution, which allow us to investigate the characteristics of the PBPs and construct non-LTE semi-empirical models. We describe the observations in Section~\ref{Observation}. The characteristics of the PBPs are described in Section~\ref{character}. The non-LTE semi-empirical models and the radiative energy estimation of the PBPs are given in Section~\ref{models}. The estimation of the average magnetic flux density in the PBPs is given in Section~\ref{flux}. Discussion and summary are given in Section~\ref{discussion}.\\

\section{Observations}
\label{Observation}

Using the Fast Imaging Solar Spectrograph (FISS) ~\citep{Chae2013} of GST/BBSO, we made spectral observations of the active region NOAA 11765 (N09E10) near the center of solar disk from 16:50 UT to 19:00 UT on 2013 June 6. FISS is a dual-band echelle spectrograph, with two cameras, one for H$\alpha$ band with 512 $\times$ 256 effective pixels, and the other for \ion{Ca}{2} 8542 {\AA} band with 502 $\times$ 250 effective pixels. Through a fast scanning of the slit, high-resolution 2D imaging spectra at the two bands can be obtained simultaneously. Each scan covered 150 steps with a spacing of $0^{\prime\prime}.16$ and lasted about 30 s. The dispersions for the H$\alpha$ and \ion{Ca}{2} 8542 {\AA} lines are 0.019 {\AA} pixel$^{-1}$ and 0.026 {\AA} pixel$^{-1}$, respectively. The spatial sampling along the slit was $0^{\prime\prime}.16 \ \rm{pixel}^{-1}$. The field of view (FOV) of each scan is about $40^{\prime\prime}\times25^{\prime\prime}$. The exposure times were 30 ms and 60 ms for H$\alpha$ and \ion{Ca}{2} 8542 {\AA} lines, respectively.

TiO images at 7057 {\AA} are recorded by the Broadband Filter Imager (BFI) with a cadence of 12 -- 13 s. The filter bandpass of TiO 7057 {\AA} observation is 10 {\AA} and the typical exposure time is 0.8 --1.5 ms. The FOV is $70^{\prime\prime}\times70^{\prime\prime}$ and the pixel size is $0^{\prime\prime}.034$. By use of the speckle reconstruction algorithm, the TiO images with high-spatial resolution were obtained. During the observations the seeing condition was better than $1^{\prime\prime}$. By use of a high-order adaptive optics (AO) systems AO-308, the diffraction limit was achieved in the observations.

\section{Characteristics of the PBPs}
\label{character}

By visual inspection of the FISS 2D spectra, three typical PBPs during the observations were detected. Table~\ref{tab1} lists some characteristics of the selected PBPs, including the time of the maximum intensity, lifetime, size and energy. The contrasts of intensity, $C$ = $ (I-I_Q )/ I_Q$, are also given in Table~\ref{tab1}, where $I$ and $I_Q$ are the average intensity of the PBPs and that of the background quiet-Sun at $\pm$ (2 -- 3) {\AA} from the center of the H$\alpha$ and \ion{Ca}{2} lines, respectively. $C_H$ and $C_{Ca}$ are the contrasts for H$\alpha$ and \ion{Ca}{2} lines, respectively. $C_{TiO}$ is the contrast at the TiO band. Due to the different resolutions between the TiO and H$\alpha$ data, the mean TiO intensity of the PBP is measured from a box of $0^{\prime\prime}.7 \times 0^{\prime\prime}.7$ around the PBP, and the intensity of the quiet-Sun is averaged over a nearby box of $3^{\prime\prime} \times 3^{\prime\prime}$.  The lifetimes are obtained by counting the slit scanning steps during which the PBP appears. The size in the $x$-direction is estimated by counting the number of pixels along the scanning direction that the PBP occupies. The size in the $y$-direction is an average of the FWHM of the brightness distribution at $\pm$ (2 -- 3) {\AA} from the center of the H$\alpha$ line.

From Table~\ref{tab1}, it can be seen that the PBPs have sizes of subarcseconds, generally less than $0^{\prime\prime}.7$. The lifetime of PBPs is less than 503 s, more or less similar to that of the photospheric granules. The contrasts of PBPs are within 0.054 -- 0.091, not obviously varying in the observed wavelengths.

The TiO data were firstly destretched to remove the residual atmospheric seeing effects. Then we aligned the H$\alpha$ wing (-1 {\AA}) images from the 2D spectra with the TiO images by the following steps. First, the TiO image was degraded to the same pixel-scale of the H$\alpha$ image. Second, we set the H$\alpha$ image as the reference image and calculate the correlation coefficients and offsets between the two images by tracing the characteristic features (e.g., sunspot umbra and penumbra regions). Next, the degraded TiO image was shifted according to the offsets to gain the largest correlation coefficient. We then re-calculated the new offsets and correlation coefficients. This process was repeated until the offsets became stable and the correlation coefficient reached a relative maximum. Finally, the offsets were used to align and recover the original TiO images. After this process, the two kinds of data are rigidly aligned and the estimated alignment error is about $0^{\prime\prime}.2$. Through a careful co-alignment of images at different wavelengths, it is found that the positions of the PBPs in H$\alpha$ and
\ion{Ca}{2} 8542 {\AA} bands correspond to the bright points in the TiO images. Figure \ref{fig1} shows the locations of the three PBPs on the 2D H$\alpha$ and TiO images. The red, yellow and blue arrows (rectangles) indicate the positions of No.1, No.2 and No.3 PBPs, respectively (Figure~\ref{fig1}a, b). We found a close spatial correspondence between the H$\alpha$ and TiO bright points,  indicated by the corresponding arrows (Figure~\ref{fig1}c--d).

Figure~\ref{fig2} shows the H$\alpha$ and \ion{Ca}{2} 8542 \AA~spectra and line profiles for the No.3 PBP. The profiles of the nearby quiet-Sun region are also shown for comparison. The intensity calibration was made by comparing our spectra compensated for the limb darkening with the disk center spectra given in \cite{Cox2000}. We took the intensity at the far wings of the lines as the continuum one. Though it is an approximate value, it does not significantly affect the contrast value, which is obtained by the ratio of the far wing intensities of the PBPs to that of the quiet-Sun. The continuum emission of the PBPs implies a significant temperature enhancement in the photosphere, compared to the quiet-Sun region. This is quite different from typical Ellerman bombs (EBs, for example, see No.2 EB in a previous study in \cite{Li2015}), which have two emission bumps near $\pm$1.0 {\AA} of the H$\alpha$ line and near $\pm$ 0.6 {\AA} of the \ion{Ca}{2} 8542 {\AA} line. This implies that PBPs and EBs are of different origins in the solar atmosphere (see also ~\cite{Rutten2013} for review). We adopt non-LTE semi-empirical models to investigate thermal and magnetic properties of the PBPs in Section~\ref{models}.

Figure~\ref{fig3} depicts temporal variations of relative intensity of the three PBPs. The relative intensity of each PBP at each time was measured as a mean value of $(I-I_Q )/ I_{max}$ at (-2.5 -- -3) {\AA} from the center of the H$\alpha$ core, where $I_{max}$ is the maximum intensity of the PBP. Figure~\ref{fig3} indicates that the intensity variation of PBPs is gradual, which gives a constraint on the modeling of PBPs, and its cause needs to be explored further.

\section{Non-LTE semi-empirical modeling of the PBPs}
\label{models}

\subsection{Computations of non-LTE semi-empirical atmospheric models}

Using both H$\alpha$ and \ion{Ca}{2} 8542 {\AA} line profiles, combined with the continuum emission at the TiO band, we can compute the semi-empirical atmospheric models of the selected PBPs. As well known, these lines are formed in a wide range of solar atmosphere, i.e., from the photosphere to the chromosphere (e.g.,~\cite{Vernazza1981}). Thus, by fitting the observational line profiles and the TiO continuum, our semi-empirical models can give the stratifications of physical quantities, such as temperature and density, etc., not only for the photosphere, but also for the chromosphere. Here we use a non-LTE computation method similar to the one described in ~\citet{Fang2006}, with an additional constraint on the contrasts in the three wavebands. We used a four-level plus continuum and a five-level plus continuum atomic model for hydrogen and calcium, respectively. Generally speaking, given a tentative temperature stratification, we solved the statistical equilibrium equation, the radiative transfer equation, the hydrostatic equilibrium and the particle conservation equations iteratively until the computed H$\alpha$, \ion{Ca}{2} 8542 {\AA} line profiles and the contrasts in the three wavebands can be well matched. The computed continuum contrasts are obtained from the formula $C$ = $ (I-I_Q )/ I_Q$, where the synthetic continuum intensity $I$ of the PBPs is computed from our semi-empirical models, and $I_Q$ is obtained from the quiet-Sun model \citep[VALC;][]{Vernazza1981}  by using the same code. Both continuum intensities cover a wide wavelength range, from 3450 {\AA} to 11000 {\AA}. Indeed, it is better to use the magnetohydrostatic equilibrium equation if there is magnetic field in the flux tubes (PBPs). Since the analyzed PBPs are near the center of solar disc and the computation is one-dimensional (1D) along the central axis of the flux tube, the hydrostatic equilibrium can still be used as an approximation. Since the convergence is considered achieved when the relative intensity error between the two iterations is less than 10$^{-7}$ and 10$^{-8}$ for hydrogen and calcium, respectively, a large number of iterations has been performed to achieve it.

The models can reproduce both the observed H$\alpha$ and \ion{Ca}{2} 8542 {\AA} line profiles, as well as the contrasts in the three wavebands. Figure~\ref{fig4} gives the temperature stratifications of the atmospheric models of the three PBPs. For comparison, the temperature stratifications for the quiet-Sun model (VALC) and the semi-empirical plage model ~\citep[P;][]{Fang2001} are also shown in Figure~\ref{fig4}. Figure~\ref{fig5} depicts both the observed and computed line profiles of the No.3 PBP. In the computation, the microturbulence velocity was taken as the same as that in the quiet-Sun model. A Gaussian macroturbulence velocity of 5 km s$^{-1}$ is adopted to convolve the computed line profiles. The fitting of the contrasts in the three wavebands is provided in Figure~\ref{fig6}. The contrast in the TiO band depends on the spatial resolution. In fact, if we take a box of $0^{\prime\prime}.5\times 0^{\prime\prime}.5$ around the PBP,  the contrast would be 0.11, as shown in Figure~\ref{fig6}. Moreover, the shape of PBP shows an irregular structure leading to a measurement error. As a compromise, we think that the fitting is acceptable. According to the noise (fluctuation) in the line profile observations, the estimated error in the intensity measurement is less than $\pm 2\%$. Considering the approximation of continuum intensity measurement at the far wings of the spectral lines, we take $\pm 4\%$ as the total error bar for the two lines.

Figure~\ref{fig7} shows the temperature stratifications as a function of the optical depth at 5000 {\AA} ($\tau_{5000}$) for our model (PBP No.3) and several existing semi-empirical models. Among them, Lagg2010 is the KG network model taken from ~\cite{Lagg2010}, Crist2017 is a BPs model taken from ~\cite{Cristaldi2017}, and SP92 is a plage model taken from ~\cite{Solanki1992}. Lagg2010 was deduced from high-resolution Stokes observation with the \ion{Fe}{1} 5250.2 {\AA} line and without introducing a magnetic filling factor. Crist2017 used  spectropolarimetric data of the \ion{Fe}{1} 6300 {\AA} line pair in small photospheric magnetic features. SP92 was deduced from Stokes {I} and {V} spectra of four \ion{C}{1} and three \ion{Fe}{2} lines. The calculation was performed by a line ratio technique in LTE. Bell2000 represents an internal model of a flux tube \citep{Bellot2000}. It used spatially averaged Stokes {I} and {V} spectra of the \ion{Fe}{1} 6301.5 and 6302.5 lines, which were observed in a facular region. It can be seen that our model is consistent with SP92 and Crist2017 in the region of log($\tau_{5000}$) $<$ 0, while the temperatures of Lagg2010 and Bell2000 are higher and lower than that of ours, respectively. Our model is consistent with the models of Lagg2010 and Crist2017 in the region of log($\tau_{5000}$) $\ge$ 0, which is similar to the quiet-Sun model VALC, while the temperatures of Bell2000 and SP92 are higher and lower than that of ours, respectively. Note that except our model, all other models are only for the photosphere since they used only photospheric lines to retrieve physical quantities.

Our non-LTE semi-empirical models reveal significant temperature enhancement in the photosphere. The temperature enhancement is about 200 -- 500 K, compared with the quiet-Sun model. The chromospheric temperature of the PBPs is higher than that of the quiet-Sun model, but comparable to the plage model presented in ~\citep{Fang2001}. It may be caused by the enhanced radiation and/or some kinds of waves from the photosphere (e.g., ~\cite{Hasan2005}). Thus, PBPs may be an important contributor to the chromospheric heating, particularly to the heating of the plage.

According to our observed continuum contrasts and semi-empirical models, the continuum spectra of all three PBPs show a rather flat wavelength dependence with a Balmer discontinuity in which the intensity at the red side of the Balmer limit (3646 \AA) is larger than that at the blue side, contrary to the case of Type-I white-light flares \citep{Hiei1982}. These characteristics are more or less similar to the Type-II white-light flares~\citep{Machado1986,Fang1995}, and the continuum emission comes mainly from negative hydrogen ions.

It is interesting to note that in our semi-empirical models the chromospheric temperature enhancement begins at a deeper layer within PBPs than in the quiet-Sun. This result is consistent with previous results of small flux tube models ~\citep{Solanki1993,Steiner2007,Hasan2005,Criscuoli2009,Criscuoli2012}.

\subsection{Radiative energy estimation of the PBPs}

As well known, the total radiative energy mainly comes from the continuum (e.g.,~\cite{Fontenla2006,Cox2000}, and thus we neglect the contribution from spectral lines. Using the semi-empirical models and the computed continuum emission of the PBPs, we can estimate the total excess radiative energy $E$ as follows:

\begin{equation}\label{eq1}
E =  \pi {D \over 2} S_{PBP} \int_{\lambda_1}^{\lambda_2} (I_\lambda - I_{Q\lambda}) \, d{\lambda} \,\,,
\end{equation}

\noindent
Considering that the relative intensity variations of PBPs are gradual, as shown in Figure~\ref{fig3}, here we assume that the mean intensity during the lifetime ($D$) of the PBP is about half of the measured peak intensity. $S_{PBP}$ is the area of the PBP, which can be estimated by the PBP size ($x \times y$) listed in Table 1. $\lambda_1$ and $\lambda_2$ are the lower and upper wavelength limits we have used in the integration of the continuum emission of the PBP. We take $\lambda_1$ = 4000 {\AA} and $\lambda_2$ = 10000 {\AA}. $I_\lambda (0)$ and $I_{Q\lambda} (0)$ are the intensity of the PBP and the quiet-Sun at the center of solar disk, respectively. They can be estimated by use of the following formula:

\begin{equation}\label{eq3}
I_\lambda (0) = I_\lambda (\theta) / (1 - u_{2} - v_{2} + u_{2} cos \theta + v_{2} cos^2 \theta)\,,
\end{equation}

\noindent
where $I_\lambda$ ($\theta$) is the computed intensity. $\theta$ is the heliocentric angle. In our case, the observed region is near the center of solar disk with $cos \theta$ = 0.9722. $u_{2}$ and $v_{2}$ are the limb darkening coefficients which can be taken from ~\cite{Cox2000}.
We can also estimate the radiation flux $F$ in continuum emission of the PBPs as follows:

\begin{equation}\label{eq4}
F  = \pi \int_{\lambda_1}^{\lambda_2} I_\lambda(0) \, d{\lambda} \,\,,
\end{equation}

\noindent

Using the formulae (1)--(3), we have estimated the total excess radiative energies $E$ and the radiation flux $F$ of the three PBPs. The results are listed in Table 1. It can be seen that the total excess radiative energy for the PBPs is about 1.4$\times$10$^{27}$ to 2$\times$10$^{27}$ ergs. The radiation flux $F$ in the visible continuum of the PBPs is about 5.5$\times$10$^{10}$ ergs cm$^{-2}$ s$^{-1}$. There is no significant difference between the three PBPs.

There exist a few estimations of PBP's radiative energy. However, our result can be compared with those from semi-empirical models of solar active features. For example, ~\cite{Fontenla2006} give semi-empirical models for different solar phenomena in the photosphere at moderate resolution. They estimated the radiation fluxes for plage and faculae being 4.21$\times$10$^{10}$ ergs cm$^{-2}$ s$^{-1}$ and 4.17$\times$10$^{10}$ ergs cm$^{-2}$ s$^{-1}$ in the same wavelength interval as that of ours, respectively. Our estimation of the PBP's radiation flux is larger than that of Fontenla's. The difference may be due to the different spatial resolutions. It is interesting to note that the mean radiation flux of solar disk is about 2$\times$10$^{10}$ ergs cm$^{-2}$ s$^{-1}$ ~\citep{Cox2000}, which is only 36$\%$ of the PBP's value.

\section{Estimation of the average magnetic flux density in the PBPs}
\label{flux}

Several authors have pointed out that the Wilson depression occurring within magnetic flux tubes can explain the observed properties of PBPs (see, e.g.,~\cite{Solanki1993,Shelyag2010,Rutten2013}). That is, due to strong magnetic field inside a thin flux tube, the horizontal pressure balance between the interior and external quiet-Sun atmosphere will cause a lower gas pressure inside the flux tube ~\citep{Rutten2001,Vogler2005,Riethmuller2014}, which reduces the opacity so that the continuum radiation from the deeper layers with a higher temperature can escape and be observed. Up to now, unfortunately, due to the limited spatial resolution of magnetic observations, we cannot observe the flux tubes directly. However, several theoretical simulations and semi-empirical modelings have been developed to interpret the physics of bright points (see, e.g.,~\cite{Solanki1993,Steiner2007,Lagg2010,Cristaldi2017}. From the point of view of observations, it is of importance to have a good knowledge of density and temperature stratifications inside the flux tubes. Fortunately, owing to advanced GST and its AO system, we can get the spectra of PBPs, which are manifestations of small flux tubes, with a very high-spatial resolution. Using these observations, we computed the non-LTE semi-empirical models of three PBPs without introducing a filling factor. Although the models are one-dimensional, they provide estimates of the temperature and number density stratifications in the PBPs. Since the observed PBPs are near the center of solar disc, we can consider that the stratifications are along the axis of small flux tubes. Thus, using the Wilson depression effect, we can self-consistently derive the average magnetic flux density in the flux tubes as follows.

Based on the assumption of a thin-tube model ~\citep{Deinzer1984,Solanki1992}, at a given height, there should be a horizontal pressure balance with the quiet-Sun atmosphere (see Figure~\ref{fig8}). We have:

\begin{equation}\label{eq5}
P_i + {B^2 \over {8 \pi}} = P_e ,
\end{equation}

\noindent
where $B$ is the average magnetic flux density in the flux tube, $P_{i,e}$ is the gas pressure insider or outside the flux tube. In this expression, the magnetic curvature terms are neglected, i.e., the force balance reduces to pressure balance, and the vertical and horizontal components of the pressure balance decouple from each other. For an ideal gas, the gas pressure can be expressed as

\begin{equation}\label{eq6}
\begin{split}
P = N k T, \\
N = (1 + A_{He}) n_H + n_e ,
\end{split}
\end{equation}

\noindent
where $n_H$ and $n_e$ are the number density of Hydrogen and electron, respectively. $N$ is the total number density, $A_{He}$ is the relative abundance of Helium atoms. We adopted $A_{He}$ = 0.1. $k$ is the Boltzmann constant, and $T$ is temperature, which should be the same in the internal and the external atmospheres at the same height because the radiative heat will exchange quickly. As our semi-empirical models give the stratifications of temperature and all number densities $N_i$, we can get the average magnetic flux density $B$ as

\begin{equation}\label{eq7}
B = \sqrt{8 \pi (N_e - N_i) k T}\,.
\end{equation}

Figure~\ref{fig9} shows $T$ vs. $N_i$ and $N_e$, which we got from the quiet-Sun model VALC, for No.1 and No.3 PBPs. The relationship for No.2 PBP is similar to that for No.1 PBP. It is clear that only in the lower photosphere $N_e \geq N_i$. This is consistent with the Wilson depression effect and we can get $B$ from the Formule~\ref{eq7}. Figure~\ref{fig10} depicts the results. It shows that the maximum value of $B$ is approximately one kilo-Gauss, and it decreases towards both higher and lower layers, reminding us of the structure of a flux tube in the intergranular lanes between the photospheric granules, as illustrated in Figure~\ref{fig8}. Note that the "real" value of $B$ may be a bit higher than what we deduced, because the internal pressure in Formule~\ref{eq5} should be an average in the flux tube, which is less than the pressure at the flux tube center adopted from our semi-empirical models.

However, the chromospheric temperature enhancement in the PBPs, as shown in Figure~\ref{fig4}, cannot be explained by the Wilson depression effect, because in these layers $N_e < N_i$, as shown in Figure~\ref{fig9}. This may be due to other mechanisms, such as MHD waves, radiation and/or magnetic reconnection. It needs to be studied further.

\section{Discussion and conclusions}
\label{discussion}

Using GST/BBSO FISS spectral data and BFI imaging, we have obtained the spectra of H$\alpha$ and \ion{Ca}{2} 8542 {\AA} lines, and the TiO 7057 {\AA} images for three selected PBPs. These spectral data and images were obtained simultaneously with high spatial-resolution, which allow us to investigate the characteristics of the PBPs and construct non-LTE semi-empirical models.

\begin{figure}
\centering
\includegraphics[width=0.9\textwidth]{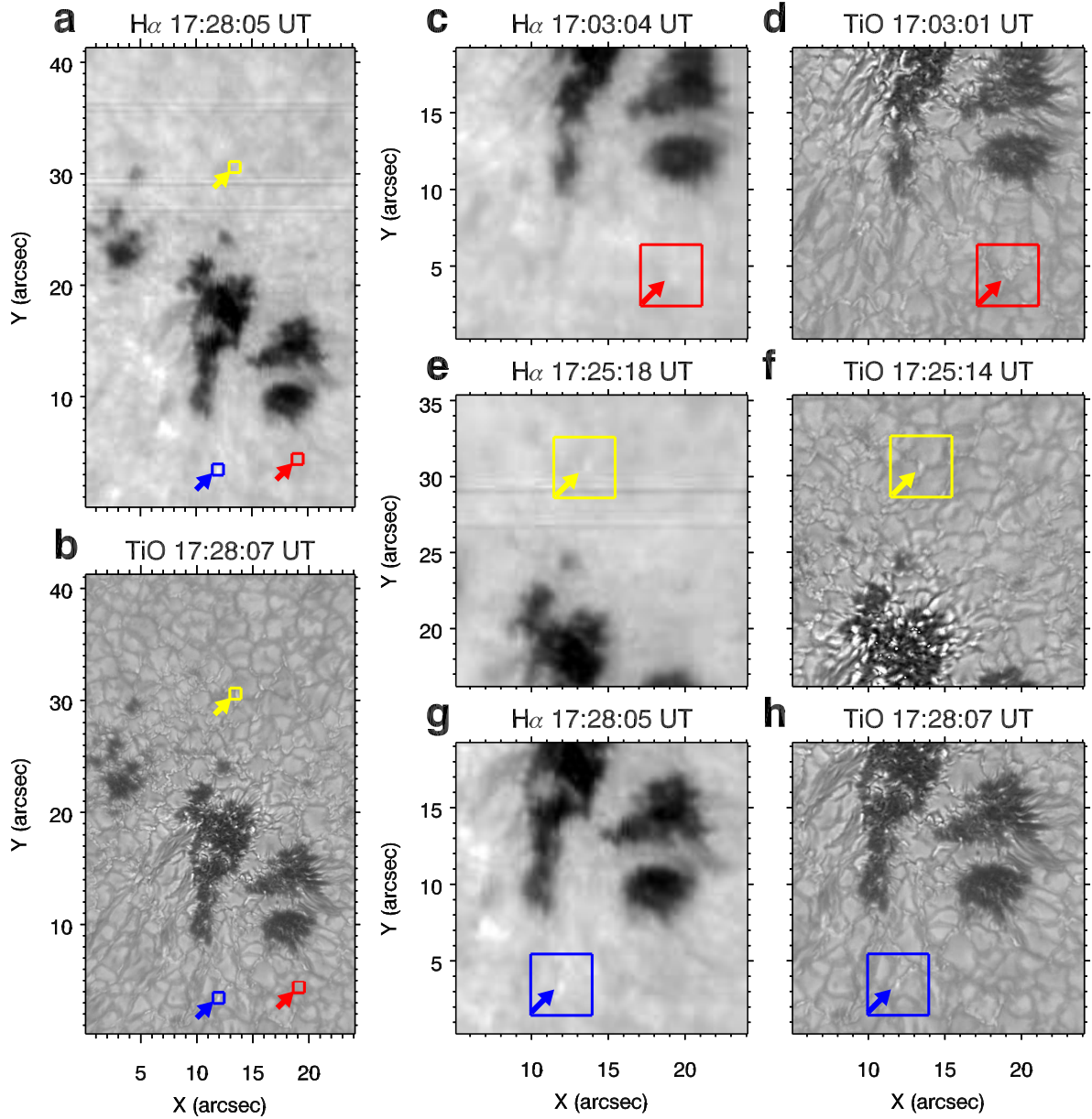}
\caption{Three PBPs observed in H$\alpha$ wing (-1 {\AA}) and TiO images. The red, yellow and blue arrows and rectangles  in (a) and (b) indicate the positions of No.1, No.2 and No.3 PBPs, respectively. (c) -- (d) Zoomed-in H$\alpha$ and TiO images showing the No.1 PBP. (e) -- (f) Similar to (c) -- (d), but for No.2 PBP. (g) -- (h) Similar to (c) -- (d), but for No.3 PBP.
} \label{fig1}
\end{figure}

\begin{figure}
\centering
\includegraphics[width=0.8\textwidth]{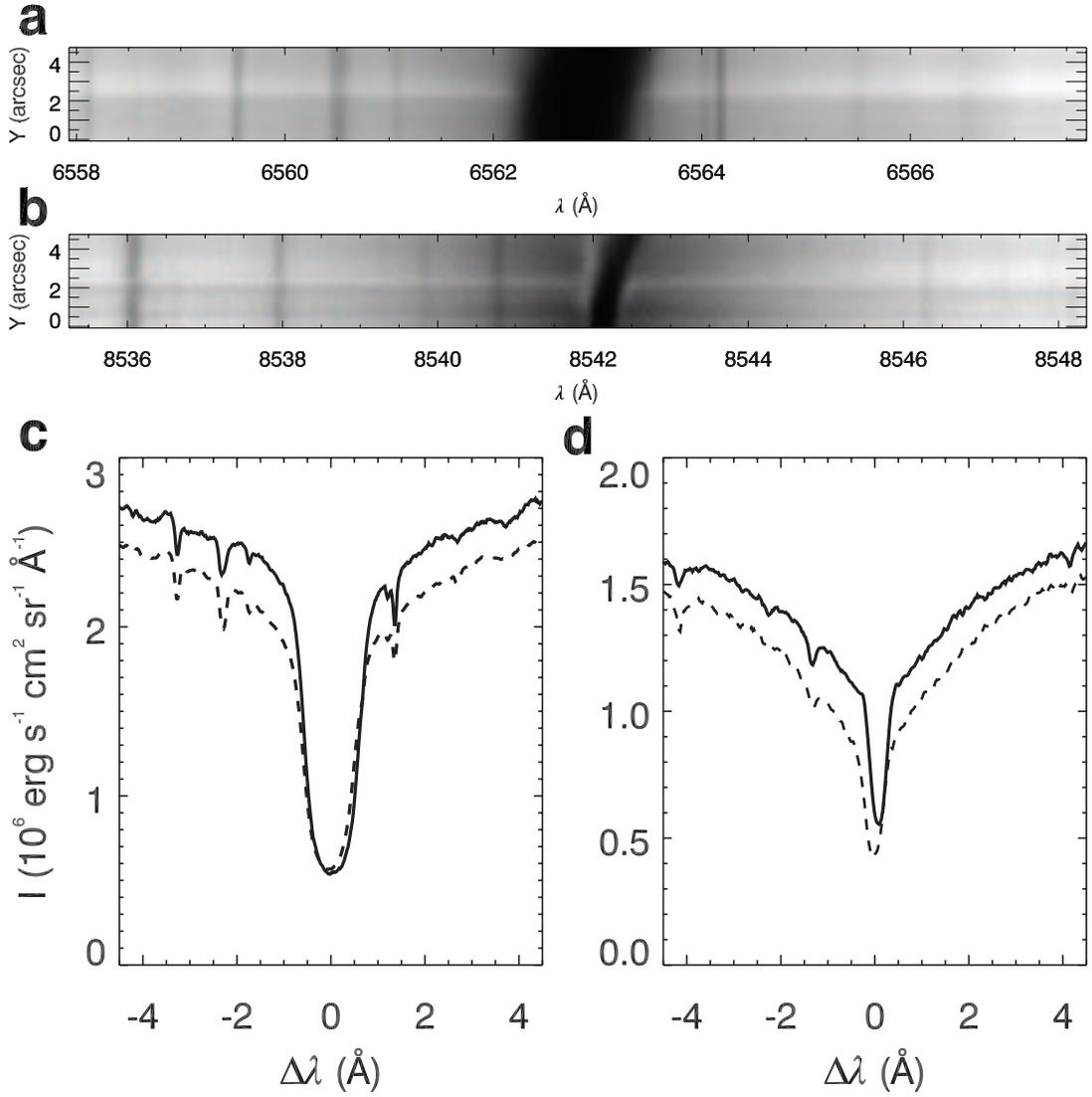}
\caption{Observed spectra and line profiles of the No.3 PBP. (a) H$\alpha$ spectrum. (b) \ion{Ca}{2} 8542 \AA\ spectrum. (c) H$\alpha$ line profiles. (d) \ion{Ca}{2} 8542 \AA\ line profiles. The line profiles for the PBP are plotted as solid lines, while the line profiles at the nearby quiet-Sun are shown as dashed lines for comparison.
} \label{fig2}
\end{figure}

\begin{figure}
\centering
\includegraphics[width=0.5\textwidth]{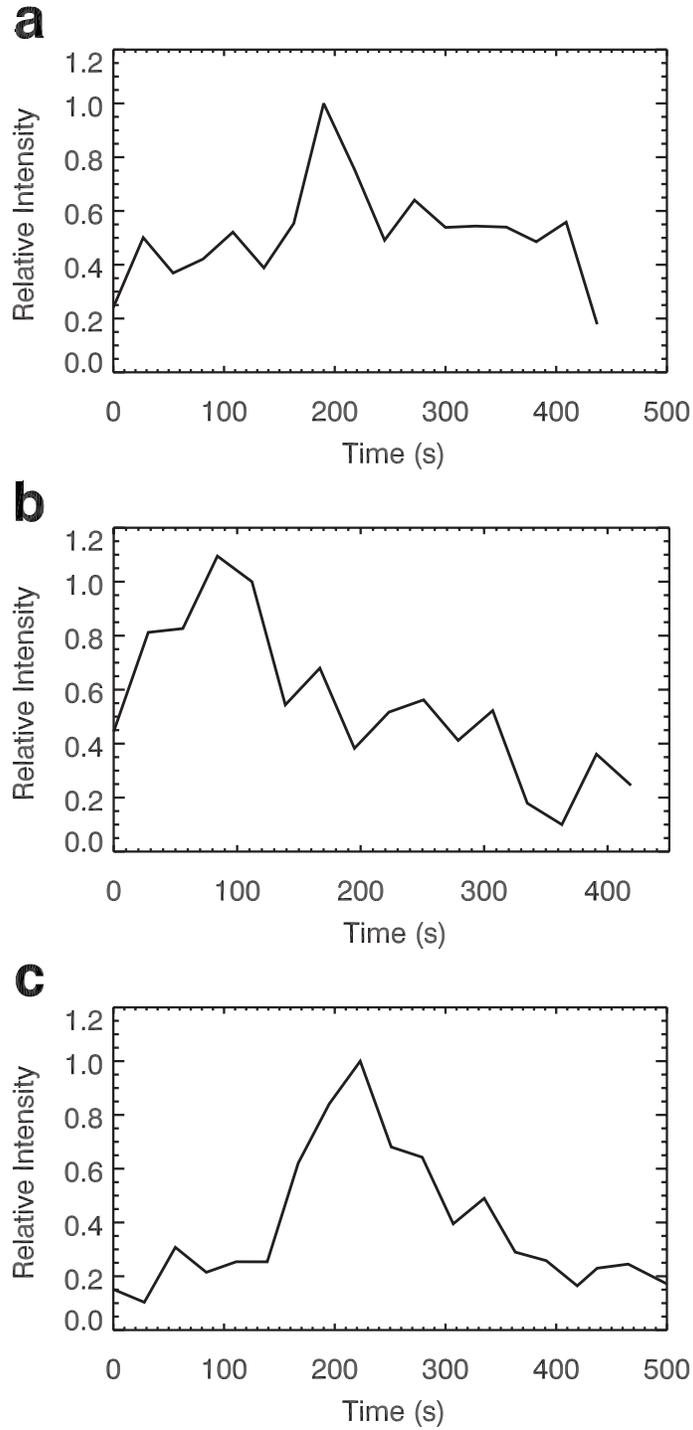}
\caption{Relative intensity variations of the three PBPs. (a), (b) and (c) correspond to No.1, No.2 and No.3 PBP, respectively. The start times for the PBPs are 16:59:54 UT, 17:23:54 UT and 17:24:22 UT, respectively. The relative intensity of each PBP at each time was measured as a mean value of $(I-I_Q )/ I_{max}$
at (-2.5 -- -3) {\AA} from the center of the H$\alpha$, where the $I_{max}$ is the maximum intensity of the PBP.} \label{fig3}
\end{figure}

The attractive property of the PBP spectra is the continuum emission component at the two far wings of both H$\alpha$ and \ion{Ca}{2} 8542 {\AA} lines (Figure~\ref{fig2}), as well as the continuum emission at the TiO band. There is one to one correspondence between the PBPs in the images at the two lines and the TiO bright points, both of which are located within the intergranular lanes between the photospheric granules (Figure~\ref{fig1}c--h).

\begin{figure}
\centering
\includegraphics[width=0.5\textwidth]{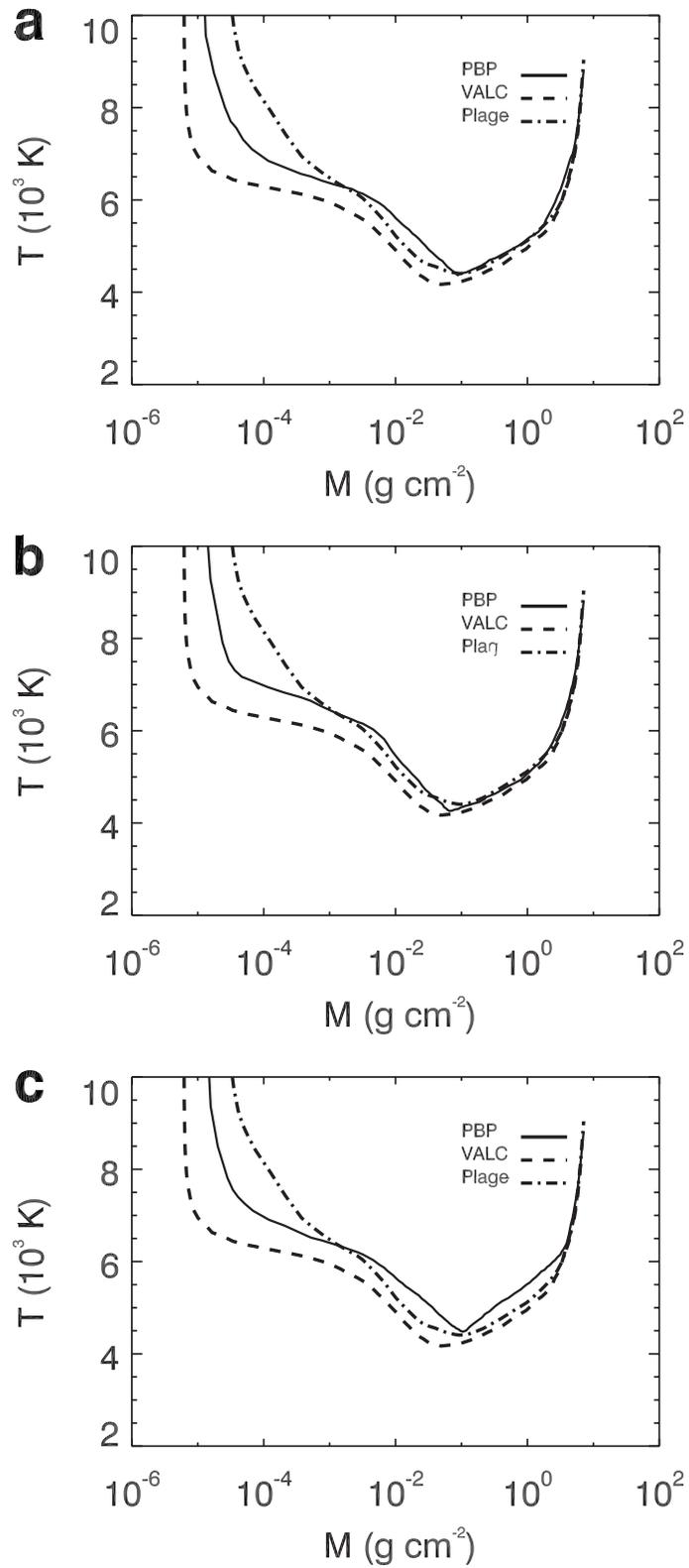}
\caption{Temperature stratifications derived for PBPs  No.1 (a), No.2 (b) and  No.3 (c) are shown as solid lines in the three panels. For comparison, the plage model described in ~\citet{Fang2001} is plotted as dash-dotted lines, and the quiet-Sun model  \citep[VALC;][]{Vernazza1981} is plotted as dashed lines in each panel.
} \label{fig4}
\end{figure}

\begin{figure}
\centering
\includegraphics[width=0.8\textwidth]{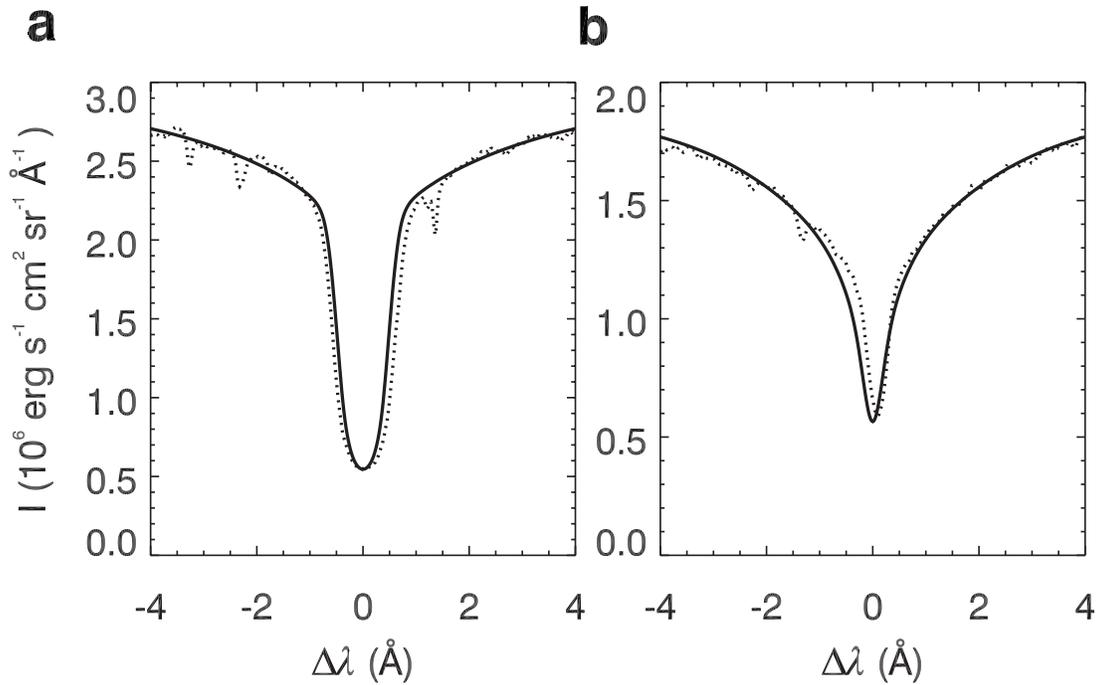}
\caption{Comparison between the observed and computed line profiles.  (a)  The observed H$\alpha$ line profile for the No.3 PBP (dotted line) in comparison with the computed line profile from the semi-empirical model (solid line).  (b)  Similar to panel (a) but for the \ion{Ca}{2} 8542 \AA\  line profiles. A Gaussian macroturbulence velocity of 5 km s$^{-1}$ is adopted to convolve the computed line profiles.
} \label{fig5}
\end{figure}

\begin{figure}
\centering
\includegraphics[width=0.8\textwidth]{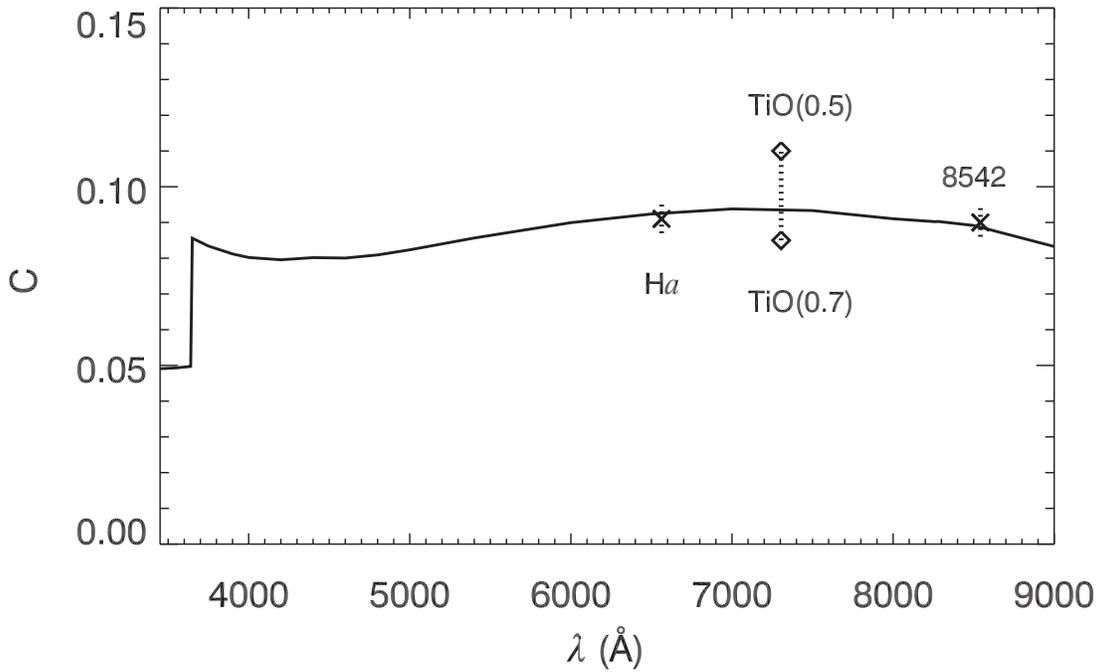}
\caption{Fitting of the contrasts in the three wavebands of No.3 PBP. The crosses refer to observations of H$\alpha$ and \ion{Ca}{2} 8542 \AA\ lines, the two rhombuses correspond to the values obtained from the TiO band in a box of $0^{\prime\prime}.7\times 0^{\prime\prime}.7$ and $0^{\prime\prime}.5\times 0^{\prime\prime}.5$ around the PBP, respectively, while the solid curve is the continuum emission computed from our semi-empirical model.}
\label{fig6}
\end{figure}

\begin{figure}
\centering
\includegraphics[width=0.8\textwidth]{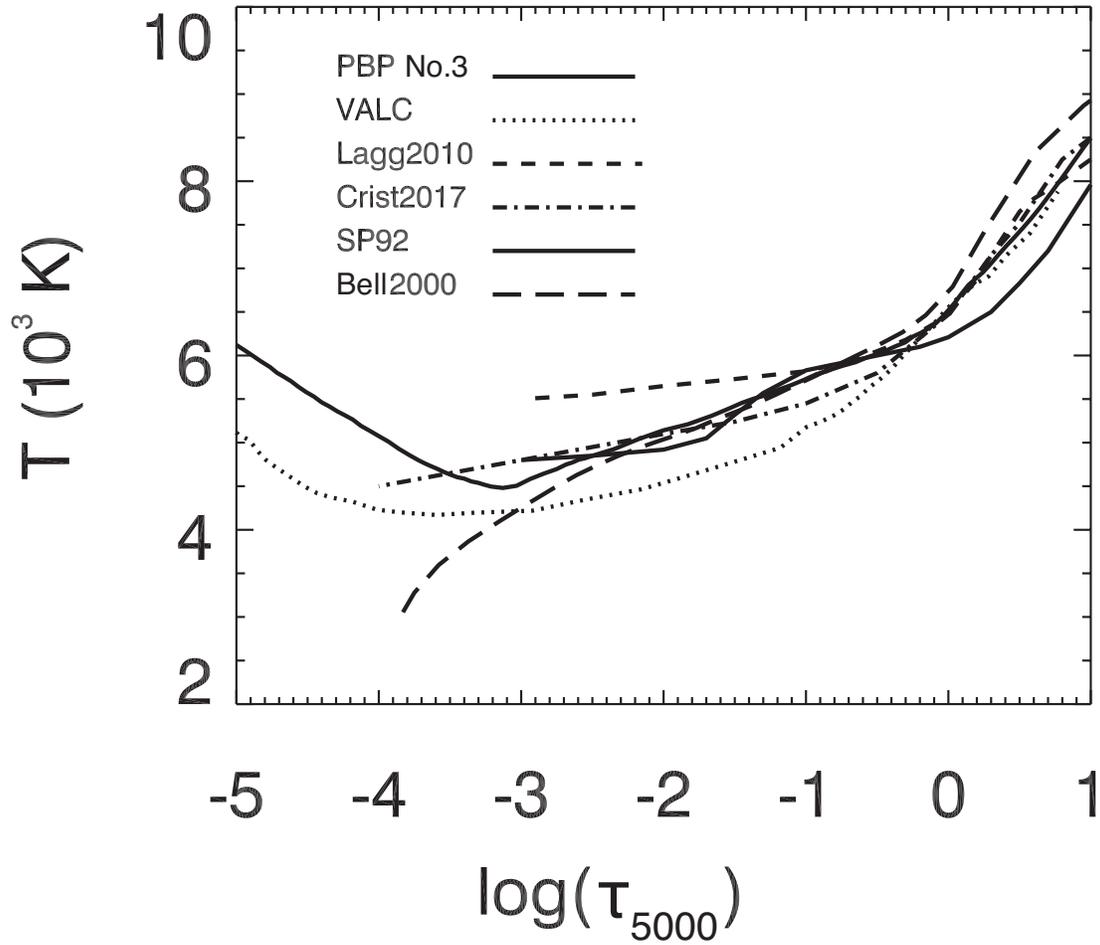}
\caption{Temperature stratifications as a function of the optical depth at 5000 {\AA} for
 the semi-empirical model of PBPs No.3 (solid line), Lagg2010 (dashed line), Crist2017 (dash-dotted line),
 SB92 (dash three-dotted line), Bell2000 (long-dashed line) and VALC (dotted line). See text for details.}
\label{fig7}
\end{figure}

\begin{figure}
\centering
\includegraphics[width=0.5\textwidth]{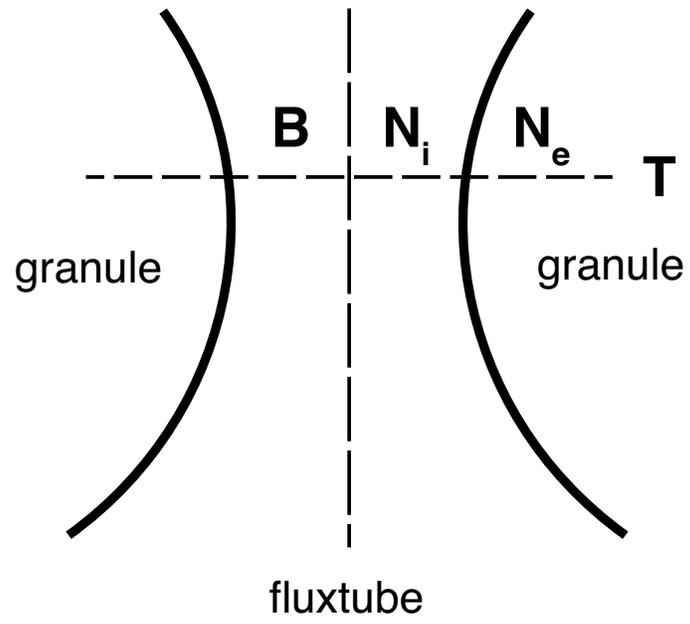}
\caption{Schematic diagram of a flux tube between two granules.}
\label{fig8}
\end{figure}

\begin{figure}
\centering
\includegraphics[width=0.5\textwidth]{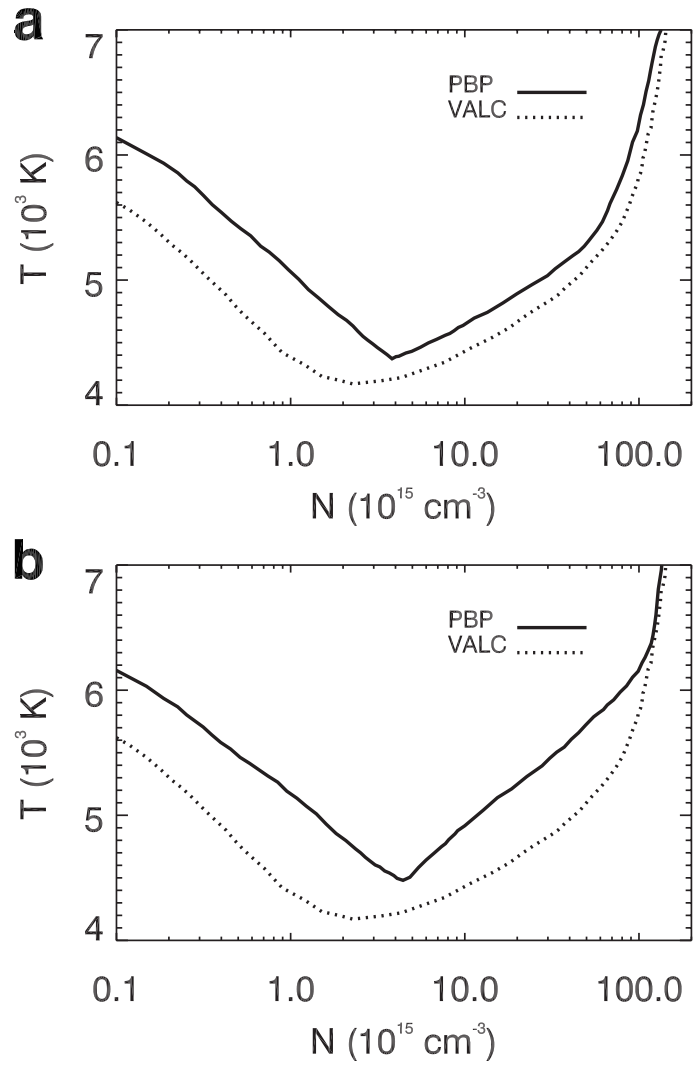}
\caption{$T$ vs. $N_{i}$ and $N_{e}$, where $N_{i}$ and $N_{e}$ are the total number density in the internal and the external atmospheres of PBPs, respectively. $N_{i}$ is taken from our semi-empirical model, and $N_{e}$ is taken from the quiet-Sun model of VALC, (a) and (b) correspond to No.1 PBP and  No.3 PBP, respectively.}
\label{fig9}
\end{figure}

\begin{figure}
\centering
\includegraphics[width=0.5\textwidth]{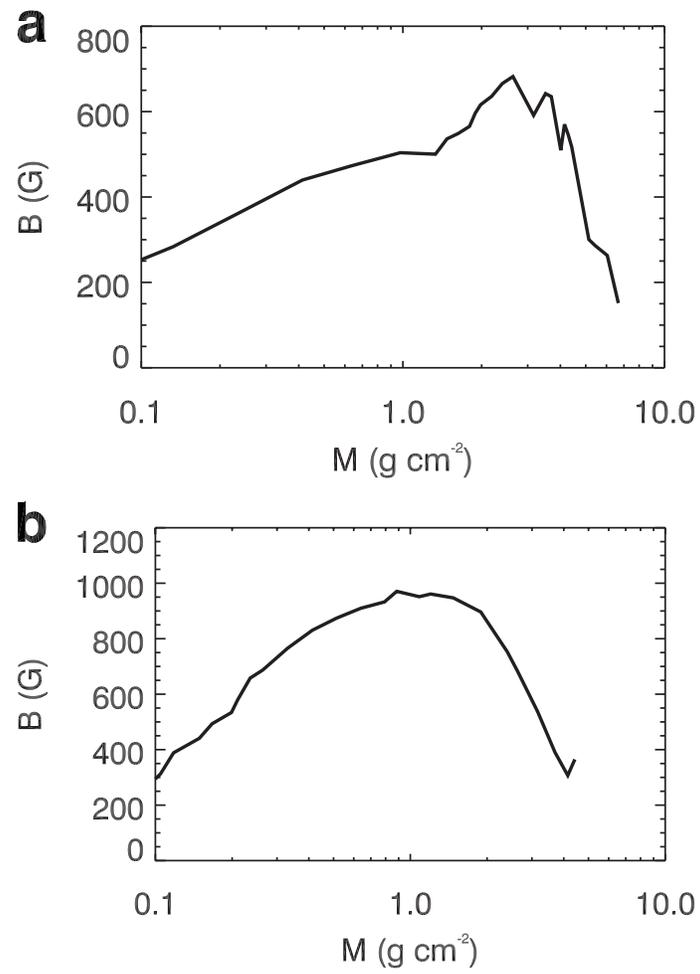}
\caption{Estimation of the average magnetic flux density $B$. (a) and (b) correspond to No.1 PBP and  No.3 PBP, respectively.}
\label{fig10}
\end{figure}

Using non-LTE theory, we have computed the semi-empirical models for the three PBPs. Our results indicate that the temperature enhancement in the lower photosphere is about 200 -- 500 K, compared with that in the quiet-Sun atmosphere at the same column mass density $M$ (Figures \ref{fig4}). By use of the observed continuum emission, combined with our semi-empirical models, the total excess radiative energy for the PBPs was estimated to be 1$\times$10$^{27}$ to 2$\times$10$^{27}$ ergs. The radiation flux $F$ in the continuum emission of the PBPs is about 5.5$\times$10$^{10}$ ergs cm$^{-2}$ s$^{-1}$.

Obviously, the total energy of PBPs should be larger than the radiative energy we estimated, because other energies, such as the magnetic and kinetic energy, are not included. If we assume that the extra energy, which can be transferred upward, is comparable to the radiative energy, then the total energy flux of PBPs can be estimated as follows. According to the statistics of PBPs, the number density of PBPs on the solar surface is about 0.25 -- 0.97 PBPs/Mm$^2$ ~\citep{Sanchez2010,Keys2011,Feng2012,Liu2018}. If we take 0.35 PBPs/Mm$^2$ as the mean PBP density, and an extra energy of 1$\times$10$^{27}$ ergs for one PBP, then the total energy density contributed by PBPs can be estimated as 3.5$\times$10$^{10}$ ergs cm$^{-2}$. If we take the lifetime of PBPs to be about 400s, then the energy flux provided by PBPs approximates 8.8$\times$10$^{7}$ ergs cm$^{-2}$ s$^{-1}$. Thus, it is reasonable to expect that the PBPs may contribute to the heating of the solar upper atmosphere via waves (like the Alfv\'en wave), radiation or other mechanisms.

Our non-LTE semi-empirical modeling confirms that the temperature enhancement of PBPs in the lower photosphere could be explained by the Wilson depression within the flux tubes (e.g.,~\cite{Rutten2001,Vogler2005,Riethmuller2014}).
Moreover, as PBPs manifest the small magnetic flux tubes, using our semi-empirical models and under the thin-tube assumption, we derived the average magnetic flux density of the flux tubes to be one kilo-Gauss.

It should be mentioned that the temperature enhancement in the PBPs obtained in the chromosphere, as shown in Figure~\ref{fig4}, cannot be explained by the Wilson depression effect. This may be due to the other mechanism, such as MHD waves, radiation and/or Joule heating. Since we do not have high resolution magnetograms, this topic is beyond the scope of this paper.

We summarize the conclusions as follows:

1. The attractive characteristics of the PBPs spectra are the continuum emission component at the far wings ($\geq$ $\pm$3 {\AA}) of both the H$\alpha$ and \ion{Ca}{2} 8542 {\AA} lines, as well as in the continuum of the TiO band. Using the continuum measurements and our semi-empirical modeling, we have estimated the radiative energy of the PBPs. Our results indicate that the total excess radiative energy of the PBPs is 1$\times$10$^{27}$ -- 2$\times$10$^{27}$ ergs, and the radiation flux in the visible continuum of the PBPs is about 5.5$\times$10$^{10}$ ergs cm$^{-2}$ s$^{-1}$. Our result reveals that the PBPs may more or less contribute to the heating of the solar upper atmosphere, in particular, to the heating of the plage.

2. The non-LTE semi-empirical atmospheric models of the PBPs have been computed. The common characteristic is the temperature enhancement in the lower photosphere. The temperature enhancement is about 200 -- 500 K. Our result also indicates that the temperature in the atmosphere above PBPs is close to the plage one. It gives a clear evidence that the PBPs may contribute significantly to the heating of the plage atmosphere. Using these models, we estimate self-consistently the average magnetic flux density $B$ in the PBPs. It is shown that the maximum value is about one kilo-Gauss, and it decreases towards both higher and lower layers, reminding us of the structure of flux tubes in the intergranular lanes between the photospheric granules.

3. We derived self-consistently the average magnetic flux density of the flux tubes to be one kilo-Gauss using our semi-empirical models with the thin-tube assumption but without introducing a filling factor.

\begin{acknowledgements}
We thank the referee for careful reading and constructive comments that helped to improve the paper. We would like to give our sincere gratitude to the staff at the Big Bear Solar Observatory (BBSO) of the New Jersey Institute of Technology (NJIT) for their enthusiastic help during CF's stay there. This work has been supported by NSFC under grants 11533005, 11203014, 11025314, 11373023, 11729301, 11703012, and 11733003. QH is also supported by the Fundamental Research Funds for the Central Universities under grant 14380032. BBSO operation is supported by NJIT and US NSF AGS-1821294 grant. GST operation is partly supported by the Korea Astronomy and Space Science Institute, the Seoul National University, the KLSA-CAS and the Operation, Maintenance and Upgrading Fund of CAS for Astronomical Telescope and Facility Instruments.

\end{acknowledgements}


\clearpage
\begin{table*}
\begin{center}
\caption{Characteristics of the PBPs}\label{tab1}
\end{center}
\begin{center}
\begin{tabular}{c c c c c c c c c c}
\hline
 No. & Time & $ C_H $ & $C_{Ca}$ &  $C_{TiO}$  & Lifetime & Size ($x\times$y) & $E$ & $F$  \\
     & (UT) &       &        &           &    (s)   & (arcsec) &  (ergs) &  (ergs cm$^{-2}$ s$^{-1}$)\\
\hline
 1 &17:03:04 & 0.089 & 0.075 & 0.079 & 437 & 0.64$\times$ 0.48 & 1.46$\times 10^{27}$ & 5.64$\times 10^{10}$ \\
 2 &17:25:18 & 0.074 & 0.063 & 0.054 & 419 & 0.64$\times$ 0.58 & 1.35$\times 10^{27}$ & 5.55$\times 10^{10}$\\
 3 &17:28:05 & 0.091 & 0.090 & 0.085 & 503 & 0.64$\times$ 0.54 & 2.03$\times 10^{27}$ & 5.67$\times 10^{10}$\\
\hline
\end{tabular}
\end{center}
\end{table*}

\clearpage

\end{document}